# Parallel Stitching of Two-Dimensional Materials

*Xi Ling[1,*], Yuxuan Lin[1,*], Qiong Ma[2], Ziqiang Wang[3], Yi Song[1], Lili Yu[1], Shengxi Huang[1], Wenjing Fang[1], Xu Zhang[1], Allen L. Hsu[1], Yaqing Bie[2], Yi-Hsien Lee[6], Yimei Zhu[5], Lijun Wu[5], Ju Li[3,4], Pablo Jarillo-Herrero[2], Mildred S. Dresselhaus[1,2], Tomás Palacios[1,‡], Jing Kong[1,‡]*

**Author Affiliations:**
[1] Department of Electrical Engineering and Computer Science, Massachusetts Institute of Technology, Cambridge, Massachusetts 02139, USA
[2] Department of Physics, Massachusetts Institute of Technology, Cambridge, Massachusetts 02139, USA
[3] Department of Materials Science and Engineering, Massachusetts Institute of Technology, Cambridge, Massachusetts 02139, USA
[4] Department of Nuclear Science and Engineering, Massachusetts Institute of Technology, Cambridge, Massachusetts 02139, USA
[5] Condensed Matter Physics and Materials Science Department, Brookhaven National Laboratory, Upton, New York 11973, USA
[6] Materials Science and Engineering, National Tsing-Hua University, Hsinchu, 20013, Taiwan

**Author Notes:**
[‡]Corresponding authors E-mail: jingkong@mit.edu; tpalacios@mit.edu
[*] These authors contributed equally to this work

As the fundamental limit of Moore's law is approaching, the global semiconductor industry is intensively looking for applications beyond CMOS electronics.[1] The atomically thin and ultra-flexible nature of two-dimensional (2D) materials (such as graphene, hexagonal boron nitride (hBN) and transition metal dichalcogenides (TMDs)) offer a competitive solution not only to push the forefront of semiconductor technology further, towards or perhaps beyond the Moore's law, but also to potentially realize a vision of ubiquitous electronics and optoelectronics in the near future.[2,3] Hybrid structures between 2D materials are essential building blocks with multi-functionality and broader capacity for nanoscale modern electronics and optoelectronics.[4–10] The stacking of van der Waals heterostructures in the vertical direction can be accomplished by either mechanical transfer or hetero-epitaxy,[4–6,8,9] whereas atomic stitching of 2D materials in the horizontal direction through conventional nanofabrication technology has proven to be far more challenging, mainly because of the lack of a selective etching method for each specific 2D material. Precise spatial control and self-limiting processes are highly desired to design and prepare lateral heterostructures. Researchers have attempted to build lateral heterostructures between materials with similar lattice structures and small lattice mismatch, such as graphene-hBN, and TMD-TMD lateral heterostructures.[4,10–16] However, parallel connection between two atomically layers with significant crystallographic dissimilarity, such as graphene-TMD or hBN-TMD lateral heterostructures, has never been achieved. Furthermore, most of these methods are not suitable for large-scale production.

Here, we developed a general synthesis methodology to achieve in-plane "parallel stitched" heterostructures between a 2D and TMD materials regardless of the lattice mismatch between materials, with large-scale production capability. This enables both multifunctional electronic/optoelectronic devices and their large scale integration. In this method, during the chemical vapor deposition (CVD) of TMDs,[16] aromatic molecules, which are used as seeds to facilitate the growth, can be "sowed" either on the substrate or on another pre-deposited 2D material, in which the "sowing" process is selective depending on the wettability of seeds and surfaces. Consequently, parallel stitched or vertically stacked[17] heterostructures between TMDs and diverse 2D materials can be achieved regardless of lattice mismatch between materials. This technique offers an opportunity to synthesize most of the basic functional building blocks for electronic and optoeletronic devices in two dimensions, including metal-semiconductor (e.g. graphene-$MoS_2$), semiconductor-semiconductor (e.g. $WS_2$-$MoS_2$), and insulator-semiconductor (e.g. hBN-$MoS_2$) heterostructures. Large-scale parallel stitched graphene-$MoS_2$ heterostructures was further investigated. Unique nanometer overlapped junctions were obtained at the parallel stitched interface, which are highly desirable both as metal-semiconductor contact and functional devices/systems, such as for use in logical integrated circuits (ICs).

Figure 1a illustrates the growth procedure of the parallel stitched heterostructure between a TMD and another 2D material through the selective "sowing" of perylene-3,4,9,10-tetracarboxylic acid tetrapotassium salt (PTAS) molecules on the growth substrate. The details are described in Method and the Supplementary Section 1. The first 2D material is transferred onto a $SiO_2$/Si substrate (growth substrate) and can be patterned by lithography and etching. PTAS molecules were deposited onto a substrate (seed reservoir) placed next to the growth substrate. At the elevated growth temperature in the presence of a carrier gas, the PTAS molecules reach the growth substrate. From our studies, it was observed that PTAS is more likely to be deposited onto a hydrophilic surface ($SiO_2$) rather than on the surfaces of 2D materials, which are mostly hydrophobic.[18,19] PTAS molecules are therefore "sowed" only in the regions without the first 2D material. This promotes the growth of the second 2D material (TMD) within the $SiO_2$ region with abundant seed molecules, whereas there is very limited TMD growth in the first 2D material region, due to the negligible amount of seed molecules. Consequently, the growth of the second 2D material only occurs in the $SiO_2$ regions, allowing the formation of parallel stitched heterostructures along the edges of the first 2D material.

Figure 1b-j show the growth results of the parallel stitched heterostructures of graphene-MoS$_2$, WS$_2$-MoS$_2$ and hBN-MoS$_2$ as the prototypes of metal-semiconductor (M-S), semiconductor-semiconductor (S-S), and insulator-semiconductor (I-S) heterostructures, respectively. Optical microscopy, atomic force microscopy (AFM) (insets in Figure 1 c, e, g and Figure S2) and spectroscopy characterizations (Figure 1h-j) were carried out. These studies reveal that high quality MoS$_2$ is grown on the exposed SiO$_2$ regions, not on top of the first 2D materials, but are well connected with them at the edges with minimal overlaps. The photoluminescence (PL) and Raman spectra are collected on and outside the first 2D materials, as shown in Figure 1h-j. The intense PL signal around 1.85 eV, and the E$_{2g}$ and A$_{1g}$ Raman modes ($\Delta\omega$=21 cm$^{-1}$) obtained in the MoS$_2$ region (red traces in Figure 1h-j) indicates the high quality of the monolayer MoS$_2$.[17] While on top of the first 2D materials (black traces in Figure 1h-j), only the Raman modes from them (monolayer graphene: G-band at 1582 cm$^{-1}$ and G'-band at 2676 cm$^{-1}$; WS$_2$: 174, 295, 322, 350 and 417 cm$^{-1}$; hBN: 1368 cm$^{-1}$) were observed. The intensity mappings of the Raman (or PL) signals from the first 2D materials and the PL from MoS$_2$ further indicate that high-quality monolayer MoS$_2$ was only grown outside the first 2D materials (Figure 1c, e, and g). We consequently conclude that sharp and well-stitched boundaries were formed at the edges of the first 2D materials, with no breaks or tears.

Since there is a large lattice mismatch (25%) between graphene and MoS$_2$, the epitaxial growth between graphene and MoS$_2$ is, in principle, more difficult than that between graphene and hBN,[12] or between two different TMDs.[14–16] High resolution transmission electron microscopy (HRTEM) was therefore carried out on the graphene-MoS$_2$ parallel stitched heterostructures for structural characterization (Figure 2a-b). Selected-area TEM diffractograms indicate that the dark area between the dashed lines in Figure 2b consists of graphene and MoS$_2$, while only MoS$_2$ (left side) or graphene (right side) can be observed outside the interface region. This indicates that MoS$_2$ overlaps with graphene at the boundary between them. The width of most of the overlapping region is 2 nm -30 nm (Figure 2b and Figure S4). Figure 2b shows a heterojunction with an overlap region of only 2.5 nm wide. The atomic structures are clearly seen on the MoS$_2$, graphene, and overlapping regions (Figure 2b and S5). Figure 2c-e show the diffractograms through fast Fourier transform (FFT) in each of these three regions. The red circle marks the diffraction pattern from the MoS$_2$ lattice structure with a spacing of 2.7 Å corresponding to the (100) planes. The orange circle marks the 2.1 Å spacing from (110) planes of graphene lattice structure. The FFT diffraction pattern of the overlapping region indicates that there is a 5° rotation angle between the MoS$_2$ and graphene lattices. Comparing Figure 2d with Figure 2c and 2e, the corresponding MoS$_2$ and graphene lattice spacing individually remain at 2.7 Å and 2.1 Å, respectively, in the overlapping region, indicating that there is no lattice distortion at their interface. As the lattice mismatch between MoS$_2$ and graphene is relatively large, the unchanged lattice constants for both of them at the interface indicate that MoS$_2$ and graphene are connected with each other presumably through van de Waals interaction. Similar analysis was done on more samples, which shows that the rotation angles between the two materials are all within a small range of 0-10° (Figure S4).

A closer look reveals more atomic defects in the overlapping region (Figure 2f) than that in the regions away from the interface. There are mainly two kinds of defects in the MoS$_2$ lattice: Mo-Mo bond defects and -S- bridge defects.[20,21] The -S- bridge defect looks like an 8-member ring in the lattice, while the Mo-Mo bond defect corresponds to a 4+8-member ring. Both defects have been observed at MoS$_2$ grain boundaries.[21] In the overlapping regions of the lateral structures, however, the $MoS_2$ defects are mainly 8-member ring defects, as marked in the Figure 2f, suggesting that the absence of seed molecules on graphene is related to a lack of sulfur for $MoS_2$ growth. Possibly, the increased density of defects at the interface results in the inhibition of $MoS_2$ growth further into the graphene region.

One unique advantage of the parallel stitching method is that it enables large scale integration. Here, we demonstrate the construction of many graphene-$MoS_2$ parallel stitched heterostructures with arbitrary patterns (Figure 3 and Figure S6). Figure 3 shows the typical optical images and spectroscopy characterization of a parallel stitched graphene-$MoS_2$ heterostructure array (Figure 3a-c), "MIT" logo (Figure 3d and Figure S6), rings (Figure 3e) and MIT mascot "Tim the beaver" (Figure 3f-h). It is observed that the mapping images of the G-band Raman intensity from graphene (Figure 3b, e and h) and PL intensity from $MoS_2$ (Figure 3c, f and i) are highly correlated with each other, and match well with the corresponding optical image (Figure 3a, d, and g). The AFM image in the inset in Figure 3a indicates that the periodic heterostructures are well connected with very narrow overlaps at the interfaces. These morphological and spectroscopic measurement results give further evidence that the $MoS_2$ and graphene are separated in space and stitched together at the edges. Using this method, one can design graphene-$MoS_2$ heterojunctions at will. As shown by the $MoS_2$ filled MIT mascot "Tim the beaver" (Figure 3f-h), the spatial resolution of the $MoS_2$ patterns in the images can be as low as 1 μm (limited by the spatial resolution of the spectrometer). The geometrical flexibility, good controllability and large-scale fabrication capability offers great opportunity for 2D hybrid multifunctional applications.

Transport measurements across graphene-MoS$_2$ parallel stitched heterojunctions indicate that a weak tunneling barrier forms at the junction. The inset in Figure 4a shows an optical image of the device. As shown in Figure 4a, both the reverse-bias current (when $V_J$<0) and the forward-bias current (when $V_J$>0) increase superlinearly with the junction voltage. This behavior is an indication that a tunneling barrier is present.[22–24] Figure 4b shows the measured barrier height ($\Phi_B$) at the junction (from temperature-dependent measurements (Figure S7)) as a function of gate voltage, the barrier height is extracted to be around 70 meV at zero gate and less than 20 meV at high positive gate voltage ($V_G$>30 V), which is similar to that for the graphene-MoS$_2$ vertically stacked heterostructure.[25] The junction resistance ($R_J=V_J/I_J$) is around 0.3 kΩ·mm when MoS$_2$ is turned on ($V_G$ = 30V). Recent works have shown that graphene-MoS$_2$ vertically stacked heterostructures, when used as contacts to MoS$_2$ channels, can lead to a much lower contact resistance than conventional metal contacts.[22,25,26] However, the vertically stacked structures would suffer from various problems when the MoS$_2$ transistors are scaled down to nanometer scale, such as (i) difficulty in alignment with a high spatial resolution, (ii) the lack of a selective etching technique, and (iii) the large parasitic impedances. In contrast, the parallel stitched junctions grown with the method presented here can serve as a self-aligned lateral Ohmic contact to the MoS$_2$ channel. Such contact has been referred to as 1D contact,[27] and has been shown to be able to address these problems very well without degrading contact quality. Implementing these 1D contacts has been a tremendous challenging task previously,[27] but can be simply realized with our selective "sowing" method here.

Using graphene-MoS$_2$ parallel stitched heterostructures as the source/drain contacts of top-gated MoS$_2$ transistors (Figure S8), we fabricated arrays of the basic building blocks of integrated circuits (shown in Figure 4c-j). Figure 4d shows the microscopic image and the transistor-level schematic of a diode-connected $MoS_2$ transistor, indicating good rectifying behavior, with an on-off current ratio on the order of $10^6$, obtained according to the *I-V* characteristics in Figure 4e. Based on the direct-coupled transistor logic (DCTL) technology,[28] which has been widely used in high-speed logic circuits with low power consumption, we successfully fabricated inverter (Figure 4f,g,h) and NAND gates (Figure 4i,j), which are a complete set of logic circuits and can, in principle, realize any 2-level combinational logics. Figure 4f plots the typical voltage transfer characteristics of an inverter, with the power supply voltage ($V_{dd}$) ranging from 3V to 6V, with a full logic swing and a symmetric and abrupt on-off transition. The voltage gain, given by $A_v = dV_{out}/dV_{in}$, as shown in the inset of Figure 4f, has a sharp peak at the medium voltage level ($V_{dd}/2$), with a value up to 7. This, together with the well-matched input-output range, guarantees the normal operation when multiple stages of logic gates are cascaded. Figure 4h, and 4j show the transient responses of an inverter and a NAND gate, respectively, in which the output voltage of the inverter is always the opposite of the input voltage, and the output voltage of the NAND gate is "low" only when both its inputs are "high". Note that the propagation delay of our logic gates still needs to be improved. This is just a proof of concept demonstration, and further optimizations will be made, such as resizing the transistor, work function matching,[28] etc., to address this issue.

In summary, as conventional lithography/selective etching is incapable for use in the large scale integration of 2D materials to achieve various junctions for future electronics and optoelectronics, in this work a universal methodology is proposed to address such challenges. By introducing the selective "sowing" of the molecules as seeds at different positions on a growth substrate during the synthesis of monolayer

TMD materials, in-plane heterojunctions of TMD with other 2D materials can be constructed. The method of heterojunction formation is effective, simple and powerful, not only offering solutions for large-scale 2D material integration, but also enabling versatile parallel stitched in-plane junctions which are unique in structure and properties, thus offering tremendous potential, as demonstrated through the example of large-scale manufacturing of parallel stitched graphene-$MoS_2$ heterostructures and the investigation of their potential applications.

**Experimental Section**

*Heterostructure growth.* In this work, graphene was prepared by either mechanical exfoliation or CVD growth, and transferred onto a cleaned 300 nm $SiO_2$/Si substrate. Other 2D materials ($WS_2$ and hBN) were prepared by mechanical exfoliation. For the parallel stitched heterostructures between $MoS_2$ and other 2D materials, PTAS molecules were used as seeds. The growth substrate (SiO2/Si wafer with the first 2D material, pre-annealed at $380^0C$ for 2 hours with 200 sccm Ar/200 sccm $H_2$ gas flow as protection) was suspended between two $SiO_2$/Si substrates with abundant PTAS as the seed reservoir, as shown in Figure S1. All of these substrates were faced down and placed in a crucible containing molybdenum oxide ($MoO_3$) precursor. This crucible was put in the middle of a 1 inch quartz tube reaction chamber, with another sulfur (S) containing crucible upstream in the quartz tube. Before heating, the whole CVD system was purged with 1000 sccm Ar (99.999% purity) for 3 min. Then, 5 sccm Ar was introduced into the system as a carrier gas. The system was heated to 650 °C at a rate of 15 °C /min, and $MoS_2$ was synthesized at 650 °C for 3 min under atmospheric pressure. The temperature at the position where the sulfur was located was set to be around 180 $^0C$ during growth. The system was finally cooled down to room temperature quickly by opening the furnace and taking out the quartz tube, and 1000 sccm Ar flow was used to remove the reactants.

*TEM characterization.* An as-grown graphene-$MoS_2$ heterostructure was transferred onto a TEM grid (Quantifoil, No.656-200-CU, TED PELLA, INC.) by the PMMA method[21]. After transfer onto a TEM grid, the PMMA was removed under vacuum annealing at 350 °C for half an hour. HRTEM characterization of the heterostructure at the atomic scale was carried out with a JOEL ARM 200CF TEM equipped with a cold field-emission electron source and two spherical-aberration correctors operated under an accelerating voltage of 80 kV to reduce radiation damage.

*AFM and spectroscopy characterization.* The AFM characterization was carried out on a Dimension 3100 instrument, commercially available from Veeco Instruments Inc. PL and Raman spectra were carried out on a Horiba Jobin-Yvon HR800 system and a Witec Alpha300-Confocal Raman Microscope. The laser excitation wavelength for the PL and Raman measurements was typically 532.5 nm. The laser power on the sample was about 0.1 mW. A 100X objective was used to focus the laser beam. The spectral parameters were obtained by fitting the peaks using Lorentzian/Gaussian mixed functions as appropriate. For Raman and PL mapping, the scan step is 0.8 μm on the Horiba Jobin-Yvon HR800 system and 0.3 μm on the Witech Alpha300-Confocal Raman Microscope.

*Device fabrication and measurement.* The $MoS_2$-graphene heterostructures were first transferred onto the 300 nm $SiO_2$/Si substrates through the PMMA transfer technique as mentioned earlier. E-beam lithography (EBL), e-beam evaporation followed by a lift-off process were used to deposit 20 nm Au as the Ohmic contacts. Another EBL and reactive ion etching (RIE) with oxygen plasma were used to define the channel area. For top-gated devices, 35 nm $Al_2O_3$ was deposited through atomic layer deposition (ALD) as the gate dielectric, followed by an e-beam evaporation of 30 nm Pd as the gate electrodes. The transport measurements were carried out in vacuum (~$10^{-5}$ Torr) using a semiconductor parameter analyzer (Agilent 4155C) and a cryogenic probe station (Lakeshore).

**Supporting Information**

Supporting Information is available from the Wiley Online Library or from the author.

**Acknowledgements**

This work was supported by the U. S. Army Research Office through the MIT Institute for Soldier Nanotechnologies, under award number 023674. X.L, Y.X.L and M.S.D acknowledge partial support from National Science Foundation under award number NSF/DMR 1004147. X.L and M.S.D acknowledge partial support from Department of Energy under award number DE-SC0001299. Y.X.L and T.P acknowledge the support from Office of Naval Research (ONR) PECASE program under award number 021302-001. Y.M.Z and L.J.W acknowledge the support from DOE-BES/MSE under Contract No. DE-AC02-98CH10886. Y.-H.L. acknowledges partial support from the Ministry of Science and Technology of the Republic of China (MOST 103-2112-M-007-001-MY3). This work was performed in part at the Center for Nanoscale Systems (CNS), a member of the National Nanotechnology Infrastructure Network (NNIN), which is supported by the National Science Foundation under NSF award no. ECS-0335765. CNS is part of Harvard University. Device fabrications were made in MIT Microsystems Technology Laboratories (MTL). TEM characterization was carried out at Brookhaven National Laboratory and MIT Center for Material Science and Engineering (CMSE). The authors thank Xiang Zhou, Xiaoting Jia, Albert D. Liao, Xiang Ji and Edbert J. Sie for their useful help.

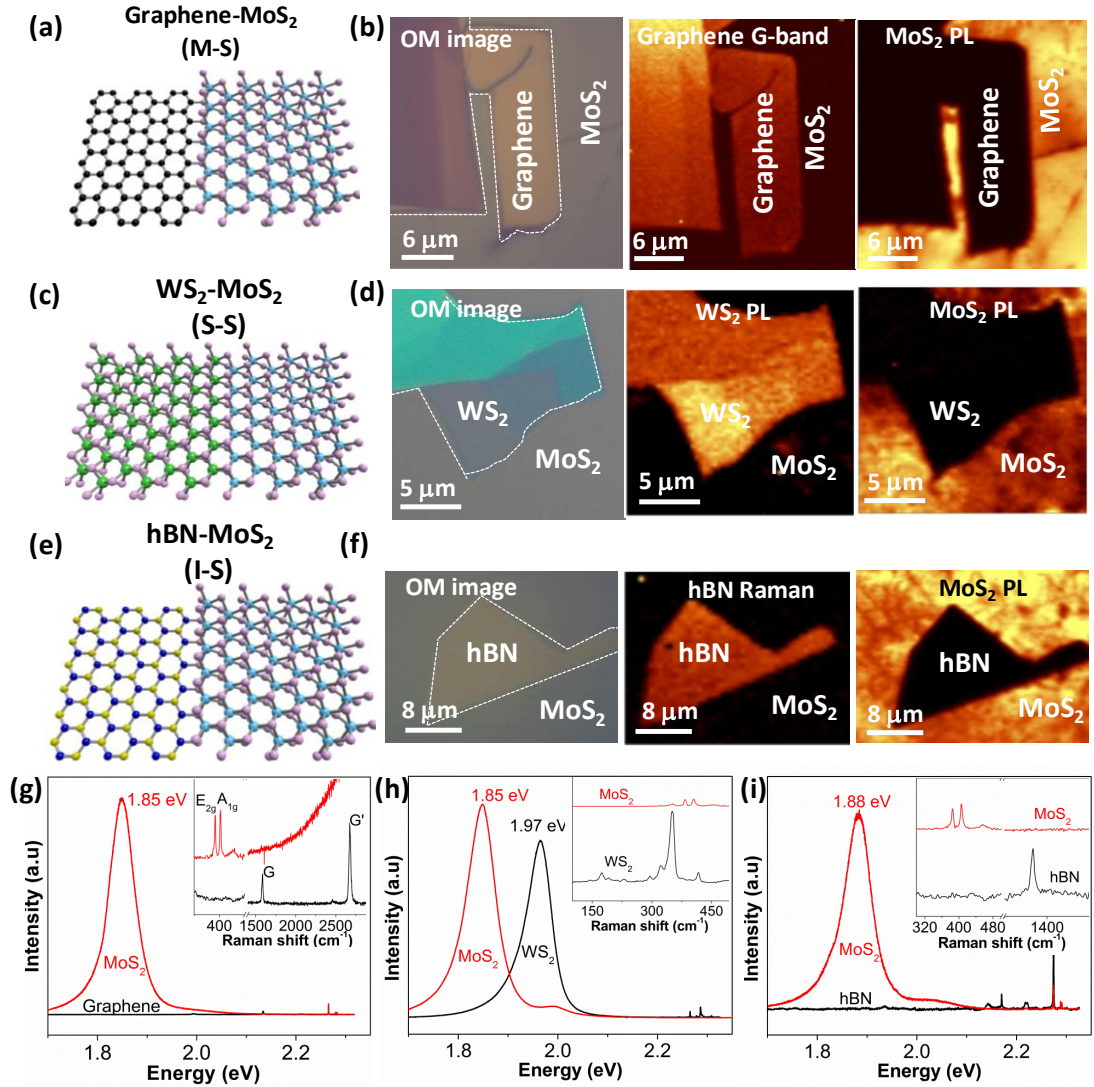

**Figure 1.** Diverse parallel stitched heterostructures between MoS$_2$ and various 2D materials. a), Schematic illustration of the CVD setup and the process for the synthesis of the parallel stitched 2D-TMD heterostructure. b,d,f), Schematic illustration of the parallel stitched heterostructures of graphene-MoS$_2$ (b), WS$_2$-MoS$_2$ (d), and hBN-MoS$_2$ (f). c,e,g), Typical optical images, spectroscopy intensity mapping images on the structures corresponding to (b), (d), and (f). The boundaries between MoS$_2$ and the pre-existing 2D materials are marked by the white dashed line. The scale bars are 5 μm. Insets in the optical images are the AFM images on the heterojunction (the scale bars are 500 nm). h,i,j), Typical PL spectra and Raman spectra (insets) collected on MoS$_2$ and the pre-existing 2D materials areas of the parallel stitched heterostructures (h), WS$_2$-MoS$_2$ (i), and hBN-MoS$_2$ (j).

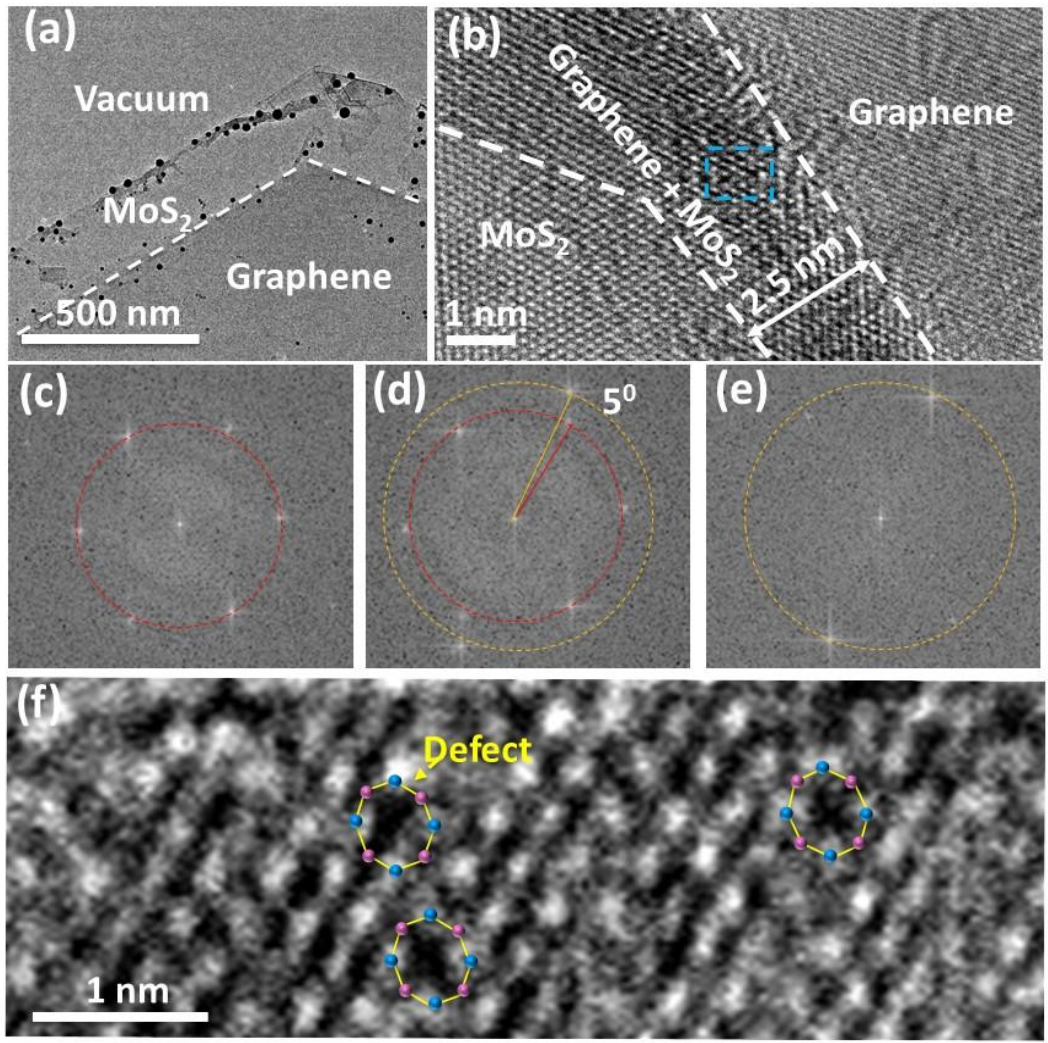

**Figure 2.** TEM characterizations of the parallel stitched graphene-MoS$_2$ heterojunction. a), Low magnification BF-TEM image showing the graphene-MoS$_2$ interface. b), HRTEM image showing the atom arrangement at the graphene-MoS$_2$ heterojunction with the size of the overlapping region of 2.5 nm. c-e), Diffractograms corresponding to the MoS$_2$ region (c), graphene-MoS$_2$ overlapping region (d) and graphene region (e), respectively. The red and orange circles mark the diffraction patterns from MoS$_2$ and graphene, respectively. f), Zoom-in HRTEM image of the graphene-MoS$_2$ overlapping area marked by blue in (b) showing the 8-member rings defects.

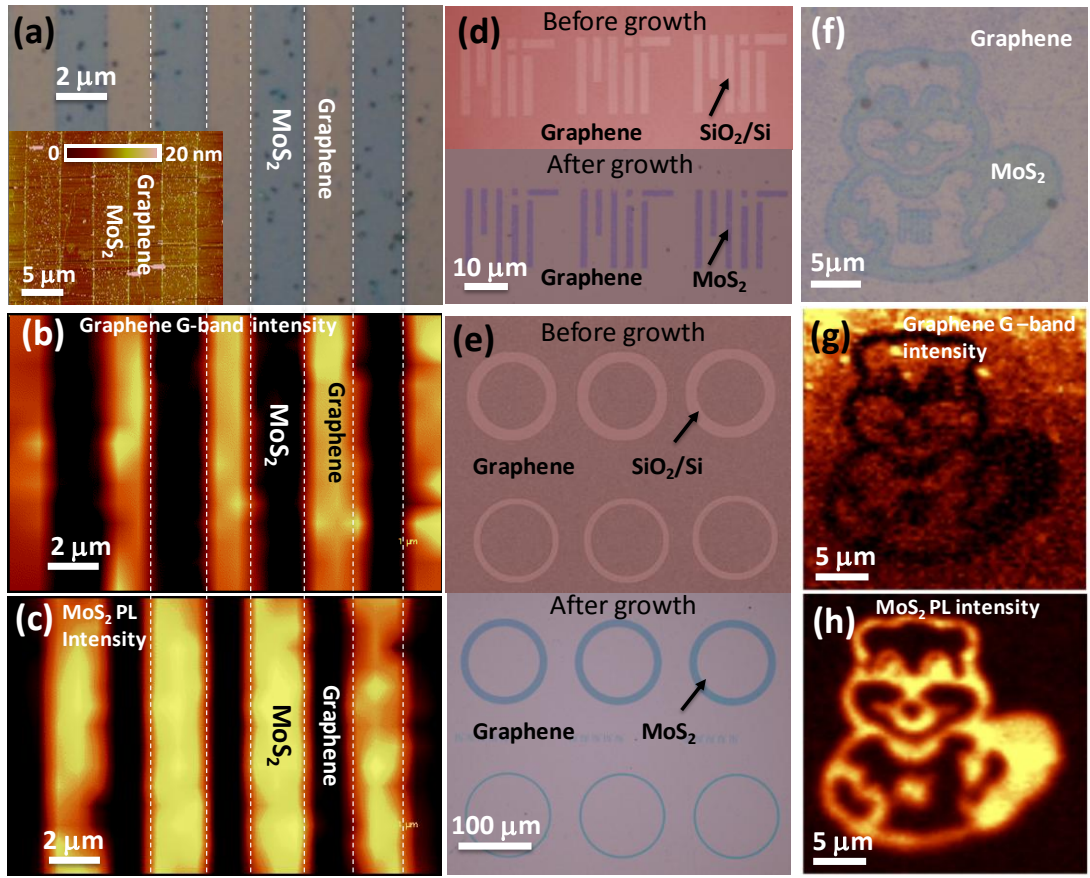

**Figure 3.** Demonstration of the parallel stitched graphene-MoS$_2$ heterojunction in a large scale with arbitrary patterns. a-c), Typical optical images of the graphene-MoS$_2$ periodic array (a), the corresponding mapping image of G-band intensity of graphene (b) and PL intensity of MoS$_2$ (c). Inset in (a) shows the typical AFM image of graphene-MoS$_2$ periodic array. d,e), Optical images before (top) and after (bottom) MoS$_2$ grown on patterned graphene pattern with "MIT" logo (d) and rings (e). f-h), Optical image of a MoS$_2$ filled MIT mascot "Tim the beaver" on graphene pattern (f), the corresponding mapping images of G-band intensity of graphene (g) and PL intensity of MoS$_2$ (h).

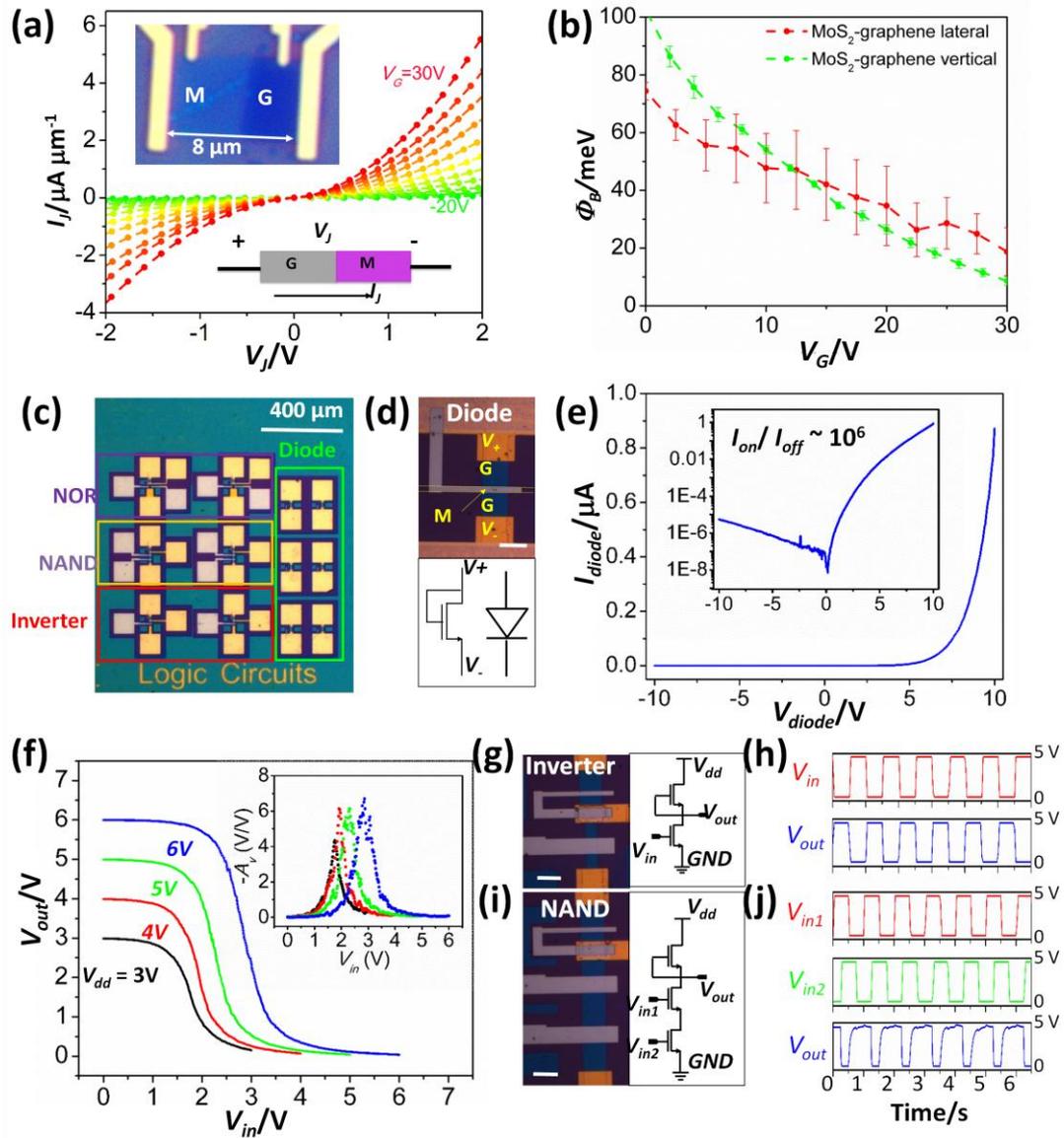

**Figure 4.** Transport measurement results and large-scale integrated circuit application of the parallel stitched graphene-MoS$_2$ heterojunction. a), Output characteristics of the graphene-MoS$_2$ heterojunction with different gate bias from -20V to 30V with 5V intervals. Upper inset: microscopic image of the 4-probe graphene-MoS$_2$ heterojunction device on top of 300 nm SiO$_2$ as the back gate dielectric. Lower inset: schematic of the device with graphene (G) as the positive side, and MoS$_2$ (M) as the negative side of the junction. b), The barrier height at the lateral (red) and vertical (green, ref. [25]) graphene-MoS$_2$ heterojunctions as a function of the gate voltage. c), Microscopic image of the test chip of the logic circuit arrays based on the parallel stitched graphene-MoS$_2$ heterojunctions. d,g,i), Microscopic images and transistor-level schematics of the diode-connected transistor (d), the inverter (g) and the NAND gate (i). Scale bar: 10 μm. E), *I-V* characteristic of the diode-connected transistor. Inset: *I-V* characteristic of the diode in a log scale. The on-off ratio is around $10^6$. F), The voltage transfer characteristic of the inverter, with the power supply voltage ($V_{dd}$)

from 3V to 6V. Inset: the corresponding voltage gain ($A_v = dV_{out}/dV_{in}$) of the inverter. h,j), transient response of the inverter (h) and the NAND gate (j).

**Parallel Stitching of Two-Dimensional Materials**

*Xi Ling[1,\*], Yuxuan Lin[1,\*], Qiong Ma[2], Ziqiang Wang[3], Yi Song[1], Lili Yu[1], Shengxi Huang[1], Wenjing Fang[1], Xu Zhang[1], Allen L. Hsu[1], Yaqing Bie[2], Yi-Hsien Lee[6], Yimei Zhu[5], Lijun Wu[5], Ju Li[3,4], Pablo Jarillo-Herrero[2], Mildred S. Dresselhaus[1,2], Tomás Palacios[1,‡], Jing Kong[1,‡]*

**Author Affiliations:**
[1] Department of Electrical Engineering and Computer Sciences, Massachusetts Institute of Technology, Cambridge, Massachusetts 02139, USA
[2] Department of Physics, Massachusetts Institute of Technology, Cambridge, Massachusetts 02139, USA
[3] Department of Materials Science and Engineering, Massachusetts Institute of Technology, Cambridge, Massachusetts 02139, USA
[4] Department of Nuclear Science and Engineering, Massachusetts Institute of Technology, Cambridge, Massachusetts 02139, USA
[5] Condensed Matter Physics and Materials Science Department, Brookhaven National Laboratory, Upton, New York 11973, USA
[6] Material Sciences and Engineering, National Tsing-Hua University, Hsinchu, 30013, Taiwan

**Author Notes:**
[‡]Corresponding authors E-mail: jingkong@mit.edu; tpalacios@mit.edu
[*] These authors contributed equally to this work

**This PDF file includes:**

**S1. Detailed description of the materials growth**

**S2. AFM characterization of the parallel stitched heterostructures between $MoS_2$ and other 2D materials**

**S3. Additional TEM characterization of graphene-$MoS_2$ parallel stitched heterostructures**

**S4. Spectroscopy characterization of the MoS$_2$ filled "MIT" logo**

**S5. Extraction of the barrier height at the parallel stitched graphene-MoS$_2$ heterojunction**

**S6. Transfer and output characteristics of top-gated MoS$_2$ field effect transistors (FETs) based on parallel stitched graphene-MoS$_2$ heterostructures**

**S1. Detailed description of the materials growth**

In this work, graphene were prepared by both mechanical exfoliation and chemical vapor deposition (CVD) growth, and transferred onto a cleaned 300 nm SiO$_2$/Si substrate. Other two-dimensional (2D) materials (WS$_2$ and hBN) were prepared by mechanical exfoliation. CVD graphene was synthesized on copper foil (99.9%, 127μm thick, Alfa Aesar) using low pressure chemical vapor deposition inside a 1 inch quartz tube. [1] Before the growth, the copper foil was annealed under 10 sccm hydrogen (H$_2$) (~400 mTor) for 30 min. During the growth, 1 sccm methane (CH$_4$) and 50 sccm H$_2$ were introduced for 1 hour (~ 1.5 mTor). After that, the copper foil was cooled down to room temperature under the same atmosphere. For the graphene transfer process, poly(methyl methacrylate) (PMMA, 4.5% in anisole) was spun coated onto the graphene/copper foil at 2500 rpm. The copper foil was etched using a copper etchant (CE-100, Transene Company Inc.). After etching, the film with graphene and PMMA was transferred to another water bath to remove the residual etchant. After replacing the water 3 times, the graphene was transferred onto a 300 nm SiO$_2$/Si substrate, and the PMMA was removed by acetone vapor.

Fig. S1a shows the schematic illustration of the CVD setup and the process for the synthesis of the parallel stitched graphene-MoS$_2$ heterostructure. To synthesize the parallel stitched heterostructure between $MoS_2$ and other 2D materials, based on the seeding promoter-assisted CVD (SPA-CVD) growth of monolayer $MoS_2$, [2–4] we chose perylene-3,4,9,10-tetracarboxylic acid tetrapotassium salt (PTAS) molecules as seeds. PTAS was prepared by mixing 3,4,9,10-perylene tetracarboxylic dianhydride (PTCDA) and ethanol and KOH and refluxing for 20 hours. PTAS was filtrated out by adding ethyl ether as an extracting agent. PTAS powder was obtained after evaporating water. 0.1 mol/L PTAS/water solution was prepared for use. Taking graphene-$MoS_2$ as an example, a 300 nm $SiO_2$/Si substrate with either exfoliated or patterned CVD graphene, which was used as a growth substrate, was suspended between two $SiO_2$/Si substrates with coated PTAS as the seed reservoir, as shown in Fig. S1a. It should be mentioned that, before loading into the furnace, annealing was done at 380 $^0$C for 2 hours to the growth substrate to remove the residue with the protection of 200 sccm Ar and 200 sccm $H_2$, which reduced the possible influence of the residues and impurities to the nucleation of the $MoS_2$. All of these substrates were faced down and placed in a crucible containing molybdenum oxide ($MoO_3$) precursor. This crucible was put in the middle of a 1 inch quartz tube reaction chamber, with another sulfur (S) containing crucible upstream in the quartz tube. Before heating, the whole CVD system was purged with 1000 sccm Ar (99.999% purity) for 3 min. Then, 5 sccm Ar was introduced into the system as a carrier gas. The system was heated to 650 °C at a rate of 15 °C /min, and $MoS_2$ was synthesized at 650 °C for 3 min under atmospheric pressure. The temperature, at the position where sulfur located, was around 180 $^0$C during growth. It should be mentioned that the average diffusion rate of S at 180 $^0$C is about 17 m/s according to the Maxwell diffusion equation, which is much larger than the 5 sccm gas flow. Therefore, the transfer of S to the substrate surrounding region was mainly by diffusion. The function of the carrier gas Ar was for diluting and controlling the concentration of S. The system was finally cooled down to room temperature quickly by opening the furnace and taking out the quartz tube, and 1000 sccm Ar flow was used to remove the reactants.

This setup allowed the PTAS seeds to be transferred from the seed reservoirs to the growth substrate under the growth temperature (650 $^0$C).[2] However, as our observation, PTAS is more likely to be deposited onto a hydrophilic surface (SiO$_2$) rather than the surfaces of 2D materials, which are highly hydrophobic, PTAS molecules are "sowed" only in the regions without graphene. Since the growth of MoS$_2$ relies on the existence of the aromatic molecular seeds in SPA-CVD, MoS$_2$ only grew on the area without graphene. Therefore, the parallel stitched graphene-MoS$_2$ heterostructure was formed with the MoS$_2$ connected to the edge of graphene (Fig. S1b). Fig. S1b shows a typical optical image of the growth results of the graphene-MoS$_2$ heterostructure, where the blue background is the monolayer MoS$_2$ film covering most of the substrate, and the flakes on the substrate are exfoliated graphene sheets with different thicknesses. The insert in Fig. S1b shows a zoom-in optical image with a MoS$_2$ monolayer grown surrounding a piece of monolayer graphene (marked by the white dashed line). Similarly, lateral heterostructures between MoS$_2$ and various 2D materials can be obtained. Fig. S1c and d show typical results for parallel stitched WS$_2$-MoS$_2$ and hBN-MoS$_2$ heterostructures, respectively.

The freedom of choice of different aromatic molecular seeds, which enhances the nucleation process in the growth, offers an opportunity to construct different types of heterostructures between MoS$_2$ and other 2D materials, because the variety of the wettability of seeds makes it possible to constrain the growth of 2D materials within a designed geometry if we locally change the wettability of the growth substrate accordingly. To synthesize the vertically stacked heterostructures, copper(II)

1,2,3,4,8,9,10,11,15,16,17,18,22,23,24,25-hexadecafluoro-29H,31H-phthalocyanine ($F_{16}$CuPc) molecules were deposited on the growth substrate (on which other 2D material was transferred beforehand) as seeds by standard vacuum thermal evaporation. Then, the regular process was carried out to synthesize monolayer $MoS_2$.[2] Since $F_{16}$CuPc could remain on graphene or hBN under the growth conditions,[2] the vertically stacked heterostructure between $MoS_2$ and graphene or hBN was obtained.

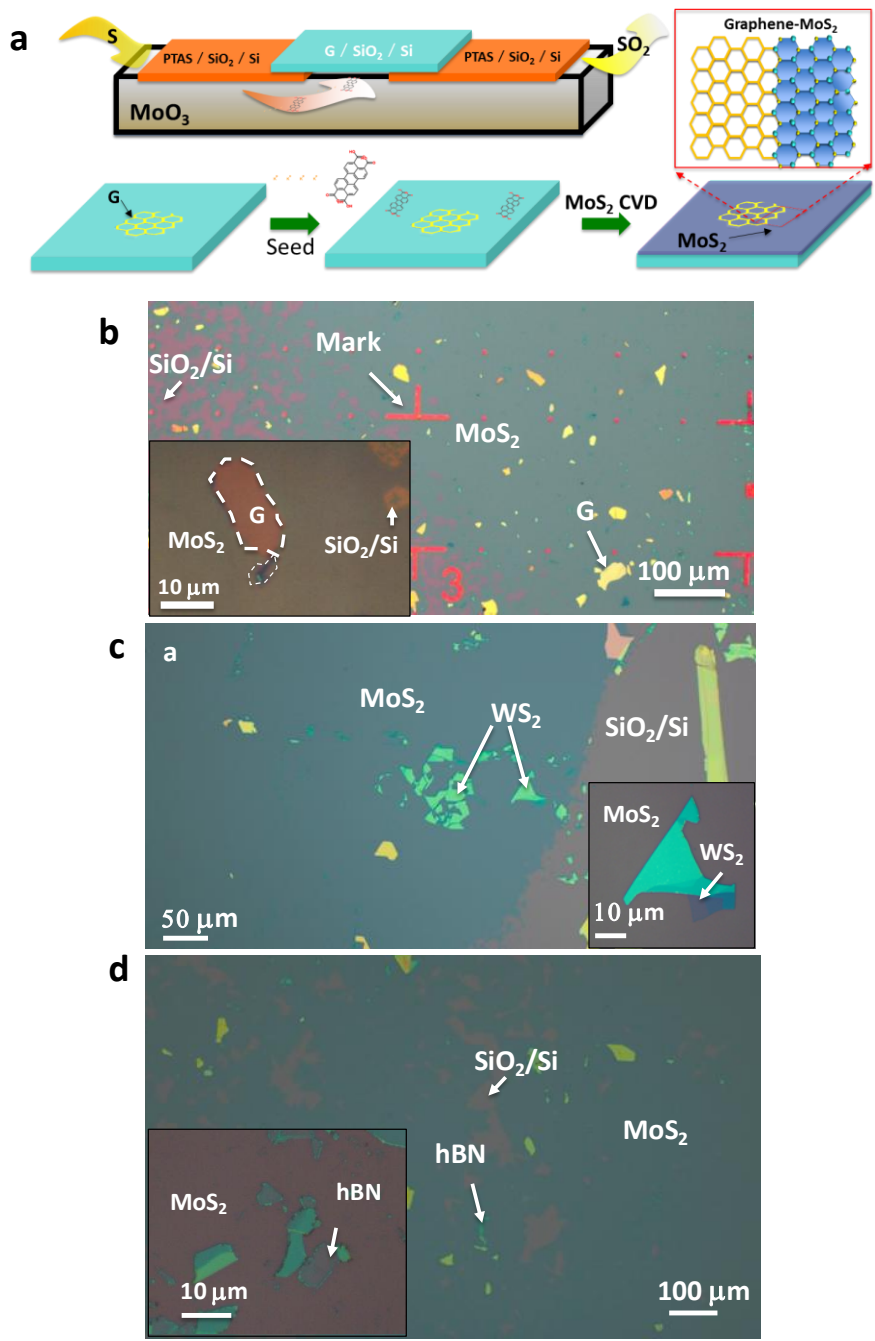

**Supplementary Figure 1 | Synthesis of the parallel stitched 2D heterostructure. a,** Schematic illustration of the CVD setup and the process for the synthesis of the parallel stitched graphene-MoS$_2$ heterostructure. **b,c,d,** Typical growth results of the parallel stitched heterostructure of graphene-MoS$_2$ **(b)**, WS$_2$-MoS$_2$ **(c)** and hBN-MoS$_2$ **(d).** The inserts show the corresponding high magnification optical images with the 2D flakes surrounded by the monolayer MoS$_2$ that was grown.

**S2. AFM characterization of the parallel stitched heterostructure between MoS$_2$ and other 2D materials**

Fig. S2 shows typical AFM images of the parallel stitched heterostructures between MoS$_2$ and diverse 2D materials (such as graphene, WS$_2$ and hBN). In all three cases, the surfaces of both the pre-existing 2D materials and MoS$_2$ are smooth and clean as shown in the AFM images. The zoom-in AFM images at the heterojunctions show the good connection between the pre-existing 2D materials and MoS$_2$. No breaks or tears are found at the junction. Fig. S3 shows the height cross-section analysis of the AFM image of a typical parallel stitched graphene-MoS$_2$ heterostructure. The boundary at the interface is 2 nm higher than the region far off the interface, which probably results from the slight overlap of MoS$_2$ on graphene and from small MoS$_2$ flakes grown on the boundary area. Since there are usually some defects on the boundary, these defects can be used as nucleation center for MoS$_2$ growth, thus we see some small triangular flakes along the boundary for some samples.

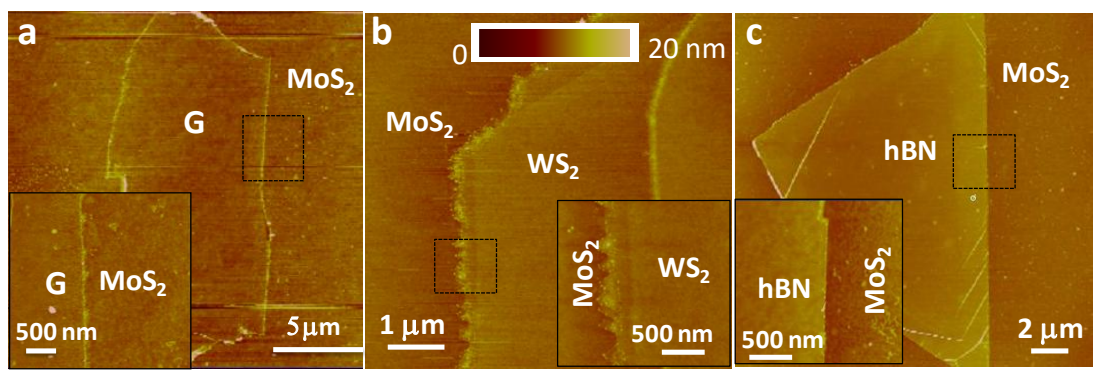

**Supplementary Figure 2 |** AFM characterization of diverse parallel stitched heterostructure between MoS$_2$ and other 2D materials. **a,b,c,** Typical AFM images of the parallel stitched graphene-MoS$_2$ heterostructure **(a)**, WS$_2$-MoS$_2$ heterostructure **(b),** and hBN-MoS$_2$ heterostructure **(c).** The inserts in each case show a zoom-in image at the heterojunction.

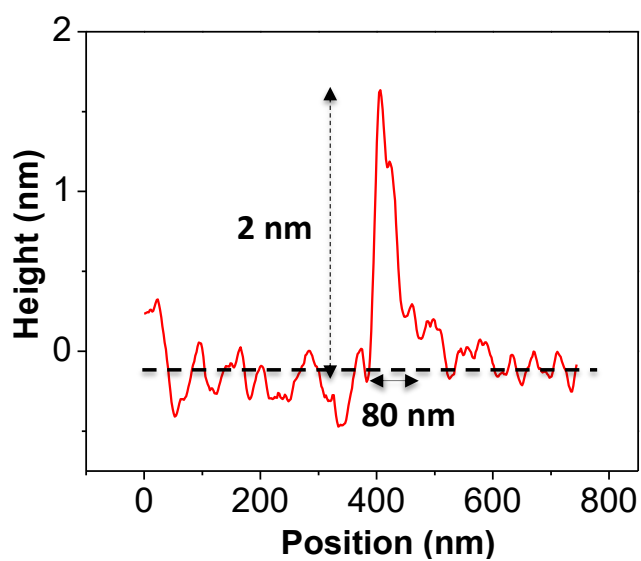

**Supplementary Figure 3 |** Height cross-section analysis of the AFM image of the overlapping region of a parallel stitched graphene-MoS$_2$ heterostructuer in Fig. S2a.

## S3. Additional TEM images for parallel stitched graphene-MoS$_2$ heterostructures

The TEM characterization of the graphene-MoS$_2$ heterostructures at different locations shows that most of the heterostructures with an overlapping region are in the several tens of nanometer size range. Fig. S4a-c shows three typical cases with overlapping sizes of 26 nm, 21 nm, and 13 nm, respectively. The corresponding diffraction patterns of the overlapping regions and the rotation angle between the graphene and MoS$_2$ regions are shown in Fig. S4d-f. Typical HRTEM images in MoS$_2$ region, graphene-MoS$_2$ overlap region, and graphene region are shown in Fig. S5. These three images in Fig. S4 show similar diffraction patterns as that in Fig. 2c-e.

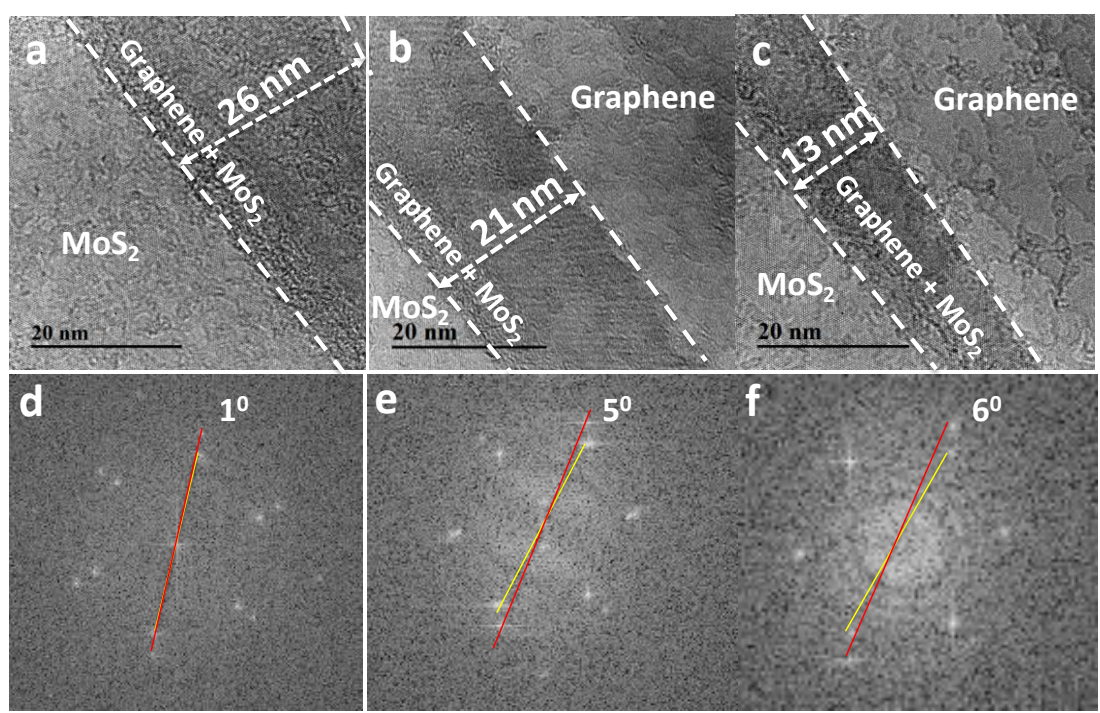

**Supplementary Figure 4 |** **High magnification BF-TEM images of the Graphene-MoS$_2$ heterostructures with different sizes for their overlapping regions. a,** 26 nm; **b,** 21 nm; **c,** 13 nm. **d,e,f,** are the corresponding diffractograms (FFT) in the overlapping areas, showing their twist angles between the two overlapped lattices.

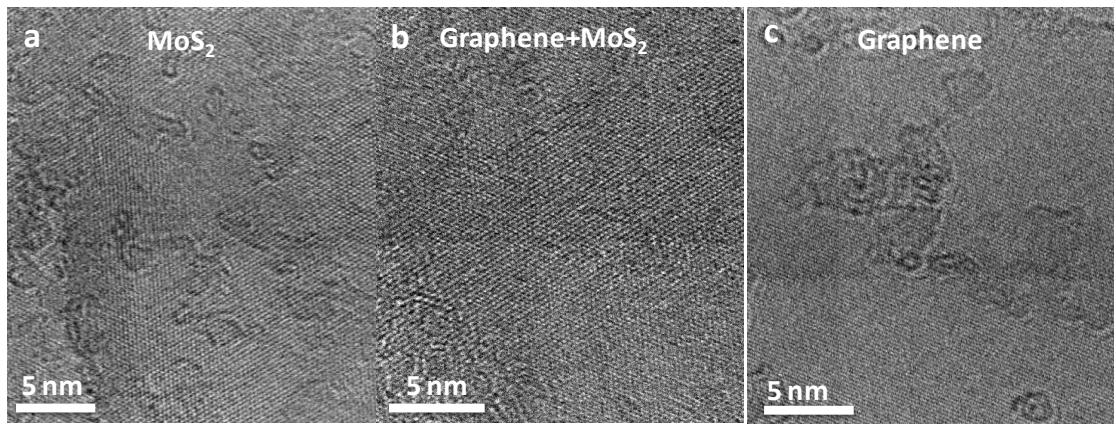

**Supplementary Figure 5 |** HRTEM images of the different areas of the parallel stitched graphene-MoS$_2$ heterostructure: **a,** the MoS$_2$ region; **b,** the graphene-MoS$_2$ overlap region; **c,** the graphene region.

## S4. Spectroscopy characterization of the MoS$_2$ filled "MIT" logo

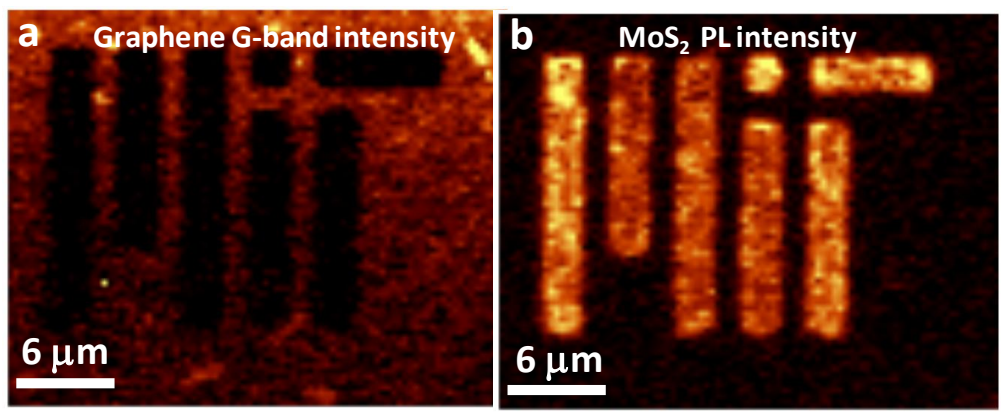

**Supplementary Figure 6 |** Mapping images of the G-band intensity of graphene (a) and PL intensity of MoS$_2$ (b) obtained on the MoS$_2$ filled "MIT" logo in Figure 3d.

## S5. Extraction of the barrier height at the parallel stitched graphene-MoS$_2$ heterojunction

Fig. S7a shows an Arrhenius plot of the junction current in the parallel stitched graphene-MoS$_2$ heterostructure with small bias ($V_J$=-0.2 V). At high temperature ($T$>100K), the junction current decreases exponentially with $1/T$; however, at low temperature ($T$<100K), a very weak temperature dependence was observed. The high-temperature regime is dominated by the combined thermionic-tunneling mechanism, and the low-temperature regime is dominated by the tunneling mechanism.

In the high-temperature region, the reverse-bias current can be expressed by:[5,6]

$$|I_R| = A^* T^{3/2} P \exp\left(-\frac{q\Phi_B}{k_B T}\right) \quad (S1)$$

where $A^*$ is the effective Richardson's constant, $k_B$ is the Boltzmann constant, q is the electron charge, P is the tunneling probability, and $\Phi_B$ is the Schottky barrier height. This equation can be rewritten as

$$\ln\left(\frac{|I_R|}{T^{3/2}}\right) = \ln(A^* P) - \frac{q\Phi_B}{k_B T}. \quad (S2)$$

The Schottky barrier height can be extracted from the slope of the linear fit between $\ln(|I_R|/T^{3/2})$ and $T^{-1}$, as shown in Fig. S7b. And the extracted barrier height as a function of the back gate voltage thus obtained is shown in Fig. 4b. The barrier height extracted from vertically stacked graphene-MoS$_2$ junctions [6] is also shown in Fig. 4b, which is very similar to that for the parallel stitched heterojunctions.

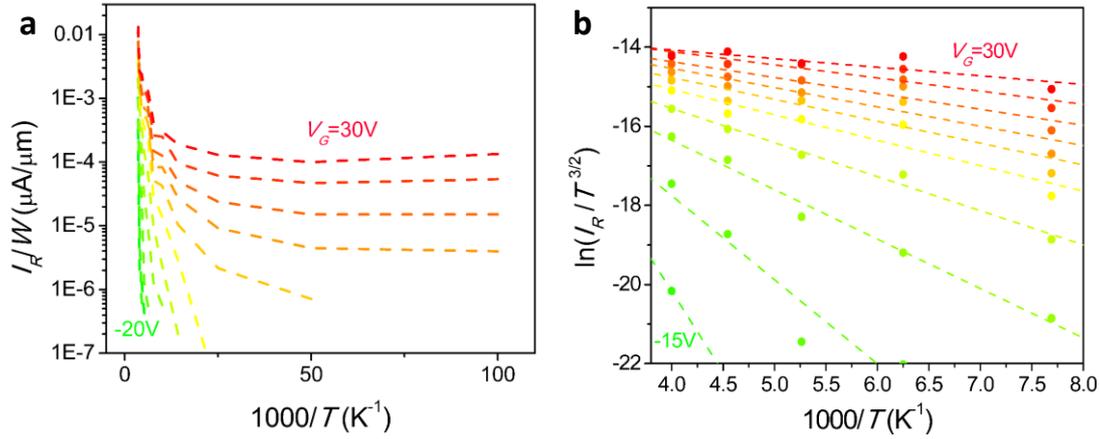

**Supplementary Figure 7 | Temperature-dependent transport measurement of the parallel stitched graphene-MoS$_2$ heterojunction. a,** The reverse-bias current density ($I_R$/W) at $V_J$=-0.2 V as a function of the reciprocal temperature (1000/T) with different gate voltages ($V_G$) ranging from -20 V to 30 V. **b,** The ln($I_R/T^{3/2}$) vs. 1000/T plot at high temperature (T>100 K). The dots are experimental data, and the dashed lines are fitted lines according to Eq. (S2).

## S6. Transfer and output characteristics of top-gated MoS$_2$ field effect transistors (FETs) based on parallel stitched graphene-MoS$_2$ heterostructures

Fig. S8a and S8b shows the transfer and output characteristics of a top-gated MoS$_2$ FET using parallel stitched graphene-MoS$_2$ heterojunctions as the source/drain contacts. Both MoS$_2$ and graphene are synthesized by CVD method. The channel length and channel width of the FET are 2 μm and 50 μm, respectively. The thickness of the Al$_2$O$_3$ top-gate dielectric film is 35 nm. Fig. S8c shows a closer view of the $I_d$-$V_d$ curves around $V_d$ = 0 V. The good linearity of the $I_d$-$V_d$ curve further proves the improved quality of the source/drain contact.

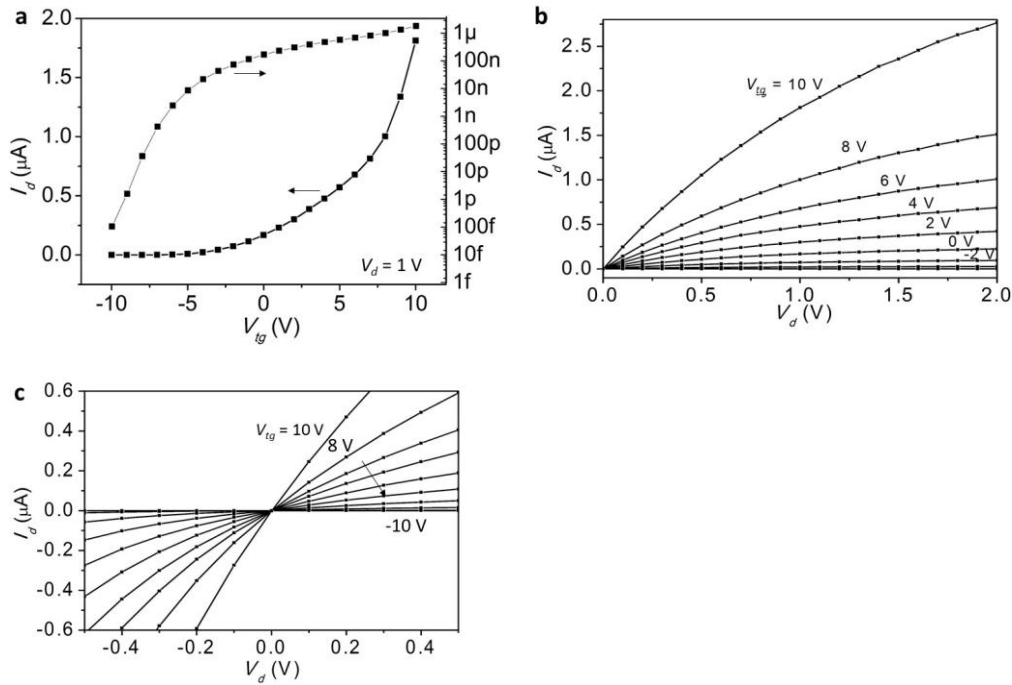

**Supplementary Figure 8 |** **Transport measurement of top-gated MoS$_2$ FETs using parallel stitched graphene-MoS$_2$ heterojunctions as the source/drain contact. a,** Transfer characteristic of the device. **b,** Output characteristic of the device. **c,** Output characteristic of the device near $V_d = 0$ V.